\documentclass[aps,prl,groupedaddress,superscriptaddress,twocolumn, longbibliography]{revtex4-1}

\usepackage[utf8x]{inputenc}
\usepackage{color}
\usepackage{bbm} 

\usepackage{amsfonts,amsmath,amssymb,stmaryrd}

\usepackage{graphicx}
\usepackage{subfigure}  
\usepackage{bbm}
\usepackage{hyperref}
\usepackage{epsfig}
\usepackage{mathrsfs}
\usepackage{verbatim}
\usepackage{ulem}	
\usepackage{siunitx}
\usepackage{soul} 

\usepackage[all]{hypcap}
\usepackage{braket}

\usepackage{array}

\DeclareSIUnit{\erec}{E_\mathrm{rec}}

\begin{document}
\normalem	

\title{Experimental realization of a Rydberg optical Feshbach resonance in a quantum many-body system}

\author{O. Thomas}
\affiliation{Department of Physics and Research Center OPTIMAS, Technische Universität Kaiserslautern, 67663 Kaiserslautern, Germany}
\affiliation{Graduate School Materials Science in Mainz, Staudinger Weg 9, 55128 Mainz, Germany}

\author{C. Lippe}
\affiliation{Department of Physics and Research Center OPTIMAS, Technische Universität Kaiserslautern, 67663 Kaiserslautern, Germany}

\author{T. Eichert}
\affiliation{Department of Physics and Research Center OPTIMAS, Technische Universität Kaiserslautern, 67663 Kaiserslautern, Germany}

\author{H. Ott*}
\affiliation{Department of Physics and Research Center OPTIMAS, Technische Universität Kaiserslautern, 67663 Kaiserslautern, Germany}

%
%

\date{\today}

\maketitle

\textbf{Feshbach resonances in ultra-cold atomic gases have led to some of the most important advances in atomic physics. They did not only enable ground breaking work in the BEC-BCS crossover regime \cite{Bloch2008RevManyBody}, but are also widely used for the association of ultra-cold molecules \cite{Chin2010RevFR}, leading to the formation of molecular Bose-Einstein condensates \cite{greiner2003emergence, Jochim2003BECMolecules} and ultra-cold dipolar molecular systems \cite{Bohn2017ColdMolecules}. Here, we demonstrate the experimental realization of an optical Feshbach resonance using ultra-long range Rydberg molecules \cite{sandor2017RydOFR}. We show their practical use by tuning the revival time of a quantum many-body system quenched into a deep optical lattice. Our results open up many applications for Rydberg optical Feshbach resonances as ultra-long range Rydberg molecules have a plenitude of available resonances for nearly all atomic species. Among the most intriguing prospects is the engineering of genuine three- and four-body interactions via coupling to trimer and tetramer molecular states \cite{Bendowsky2010Trimer}.}



Magnetic Feshbach resonances are a powerful and well established tool to change the interaction of an atomic species \cite{Chin2010RevFR}. However, they often require ultrastable and strong magnetic fields and are only available for certain atomic species and magnetic sublevels. In addition, magnetic Feshbach resonances are limited with respect to their spatial and temporal modulation. The latter can be overcome by using an optically switched magnetic Feshbach resonance \cite{bauer2009control}. An alternative approach is provided by purely optical Feshbach resonances (OFR) \cite{Nicholson2015OFR} which use a light field to couple a colliding atomic pair to a bound molecular state. While an OFR offers great spatial and temporal modulation capabilities, they typically introduce strong losses through the decay of the excited state. First experiments were therefore limited to time scales below \SI{100}{\us} \cite{Thalhammer2005OFRRb}. Experiments on millisecond time scales have become possible using dipole forbidden intercombination lines, which possess long lifetimes, but which are only available in certain atomic species \cite{Yan2013OFRintercombSr, Blatt2011OFRIdealGas, Enomoto2008OFRintercombinationYb}.

Recently, ultra-long range Rydberg molecules have been proposed as candidates for optical Feshbach resonances \cite{sandor2017RydOFR}. These molecular states are formed by the contact interaction of a quasi free electron in a Rydberg state with a ground state atom \cite{bendkowsky2009observation}. Ultra-long range Rydberg molecules are available for many atomic species and possess long lifetimes which are inherited from the underlying atomic Rydberg excitation. To our knowledge such molecules have been found in all ultra-cold Rydberg systems that have been studied so far.

\begin{figure}
	\includegraphics[width=1\columnwidth]{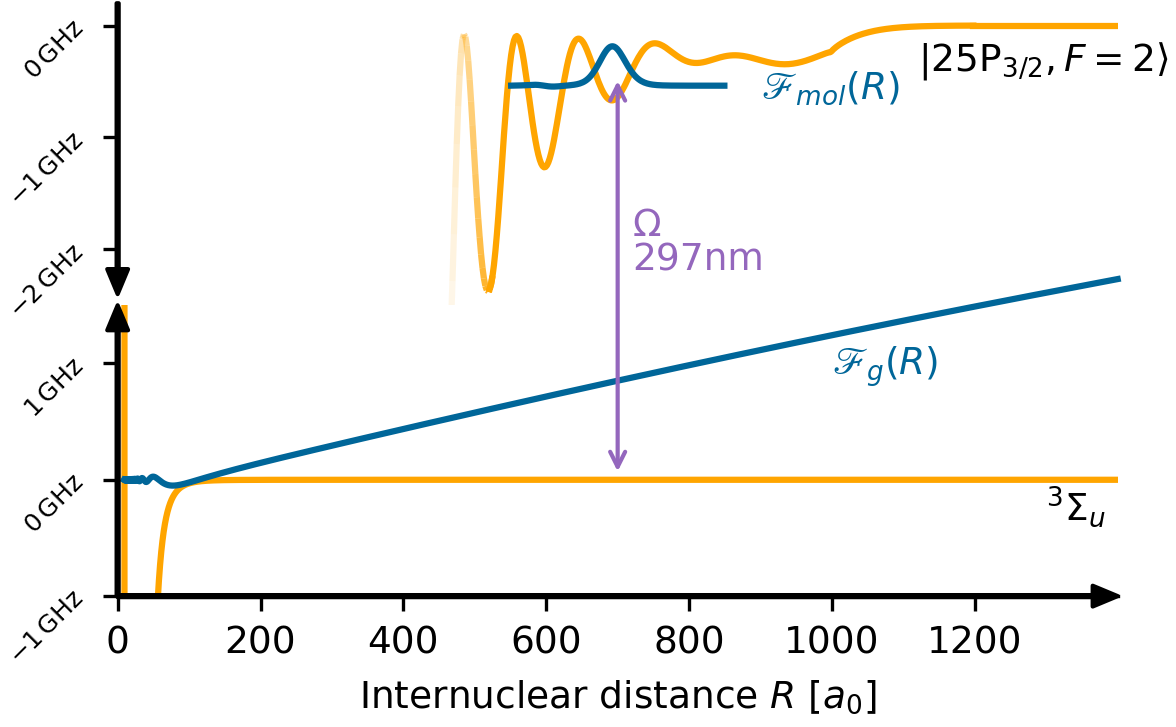}
	\caption{\label{Fig1}Schematics of the Rydberg optical Feshbach resonance. Two $^{87}\mathrm{Rb}$ atoms trapped in a 3D optical lattice scatter in the $^3\Sigma_u$ channel and are coupled with Rabi frequency $\Omega$ to an ultra-long range Rydberg molecule in the so-called "deep" potential, which is adiabatically connected to the $\ket{25\mathrm{P}_{3/2},F=2}$ state. The blue lines denote the relative wave function of the two ground state atoms ($F_g(R)$) and the molecular wave function ($F_{mol}(R)$). Due to the large size of the Rydberg molecule, the coupling takes place at a very large internuclear distances, making it independent of the short range physics of the open channel (small oscillations of $F_g(R)$ for $R<100 a_0$).}
\end{figure}

The short range molecular potential between two ground state atoms gives rise to a phase shift in the relative wave function upon a collision. At ultralow temperature, this can be cast in a single parameter, the s-wave scattering length $a$. In an OFR, the two incoming atoms (open channel) are coupled to a bound molecular state (closed channel) by a laser field. This leads to a distortion of the wave function leading to a change in the scattering length and thus a change in the interaction of the two atoms. Depending on the detuning from the molecular resonance, the scattering length can be increased or decreased. The basic principle is sketched in Fig.\,\ref{Fig1}, where we show the relative wave function of two ground state atoms and the molecular wave function. One peculiarity of a Rydberg OFR is the separation of length scales: the interatomic distance at which the molecule is formed exceeds by far the short range molecular potential of two ground state atoms. This has two consequences: first, the coupling strength of the Rydberg OFR is independent of the microscopic details of the open channel and second, the Franck-Condon overlap with the molecular state is much larger compared to previously studied optical Feshbach resonances.

In this work we specifically investigate $^{87}\mathrm{Rb}$ atoms colliding in the $^3\Sigma_u$ ground state potential which is optically coupled to the vibrational ground state of a molecular state at a bond length of $\approx\SI{700}{a_0}$ (Bohr radius) of the so-called "deep" ultra-long range Rydberg molecule potential, which adiabatically connects to the $\ket{25\mathrm{P}_{3/2}, F=2}$ state. The binding energy of the molecular state is about $h\times \SI{500}{\MHz}$. For details of the calculation and classification of the molecular potential curves see Ref.\,\cite{Anderson2014angularmomentumcoupling, Niederpr2016spinFlip}. To characterize the change in interaction strength we investigate the collapse and revival of the matter wave field of a Bose-Einstein condensate trapped in an optical lattice potential \cite{greiner2002collapse}. In the experiment, we start with a superfluid condensate of rubidium atoms in a 3D optical lattice. We then quench the optical lattice deep in the Mott insulating regime, such that the tunneling is frozen out. The matter wave field then undergoes collapse and revival dynamics, where the contrast of the interference pattern is restored after the characteristic time $T=h/U$, where $U$ is the on-site interaction energy in the lattice. The details of the experimental setup can be found in Ref.\,\cite{manthey2014SEM} and a thorough description of the experimental sequence is given in the Methods.

\begin{figure}
	\includegraphics[width=1\columnwidth]{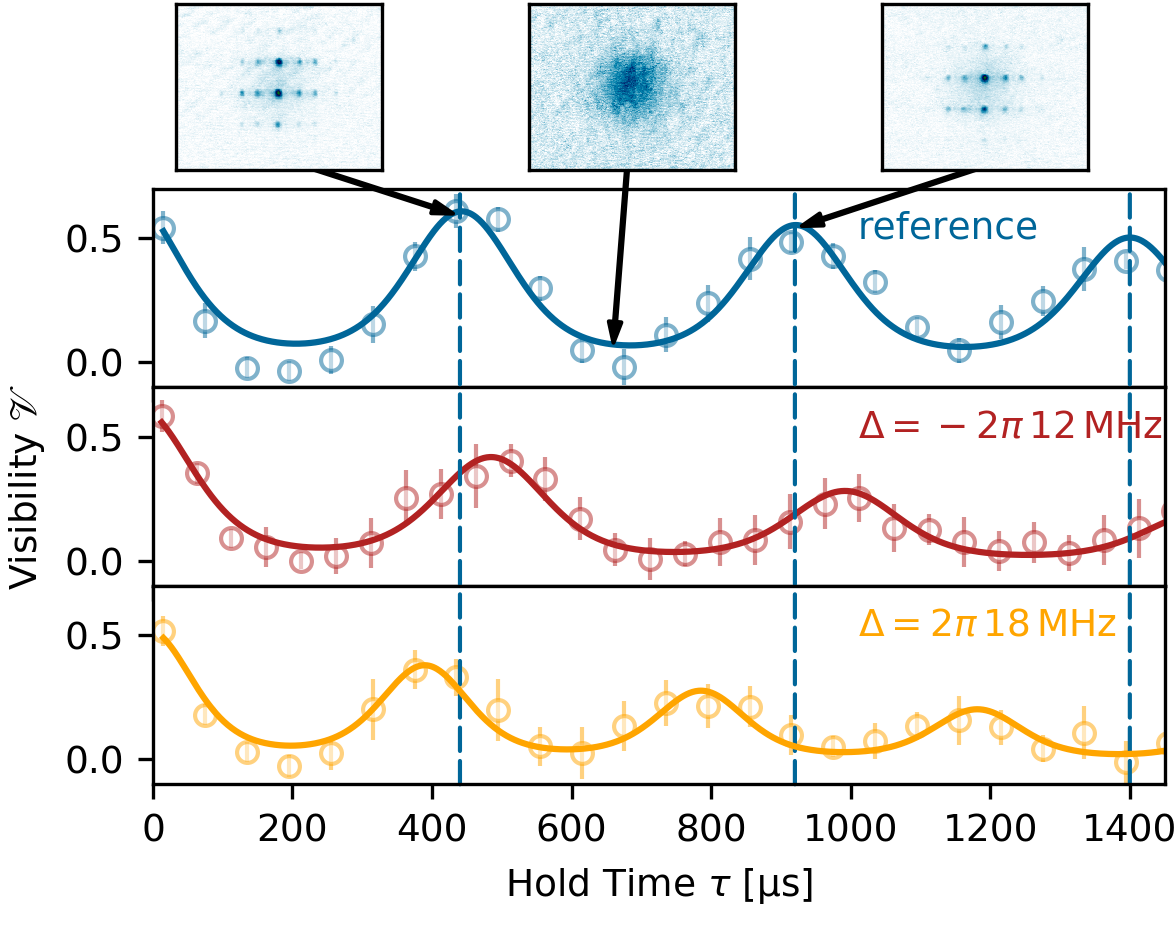}
	\caption{\label{Fig2}Collapse and revival measurements. After a variable hold time $\tau$ in the optical lattice, time-of-flight images of the cloud are taken (top row) and the super-fluid visibility is extracted (see methods). The blue data correspond to an uncoupled system, the red data to a detuning of $\Delta = \SI{-12}{\MHz}$ and the yellow data to  $\Delta = \SI{18}{\MHz}$ (both with atomic Rabi frequency $2\pi\times\SI{810}{\kHz}$). If not otherwise stated all measurements are taken with a final lattice depth of \SI{30}{\erec} (recoil energy). An increased (decreased) oscillation frequency and thus interaction strength can be seen for blue (red) detuning. For each hold time we typically record six independent measurements. The error bars donate the statistical error from averaging individual runs.}
\end{figure}

In Fig.\,\ref{Fig2}, we show the time evolution of the visibility $\mathscr{V}$ of the interference pattern after the quench. From a fit with a model function we extract the corresponding revival time and hence the on-site interaction $U$ (see methods). The upper graph shows the collapse and revival in the absence of any coupling laser and serves as a reference, $U_\mathrm{ref}$. We repeat the experiment, coupling the atoms to the molecular Rydberg state during the collapse and revival dynamics with a single photon transition at \SI{297}{\nm} (\SI{50}{\mW} power, \SI{200}{\um} waist). In Fig.\,\ref{Fig2}, it is clearly visible that the revival time can be enlarged for red detuning (red curve) and reduced for blue detuning (yellow curve). The interaction shift is then extracted as $\Delta U = U - U_\mathrm{ref}$. Varying the detuning or the laser intensity, we map out the Rydberg OFR. Our measurements show that the OFR preserves the coherence during the collapse and revival of the many-body wave functions and is therefore a direct demonstration of interaction tuning in a quantum many-body system.

\begin{figure}
	\includegraphics[width=1\columnwidth]{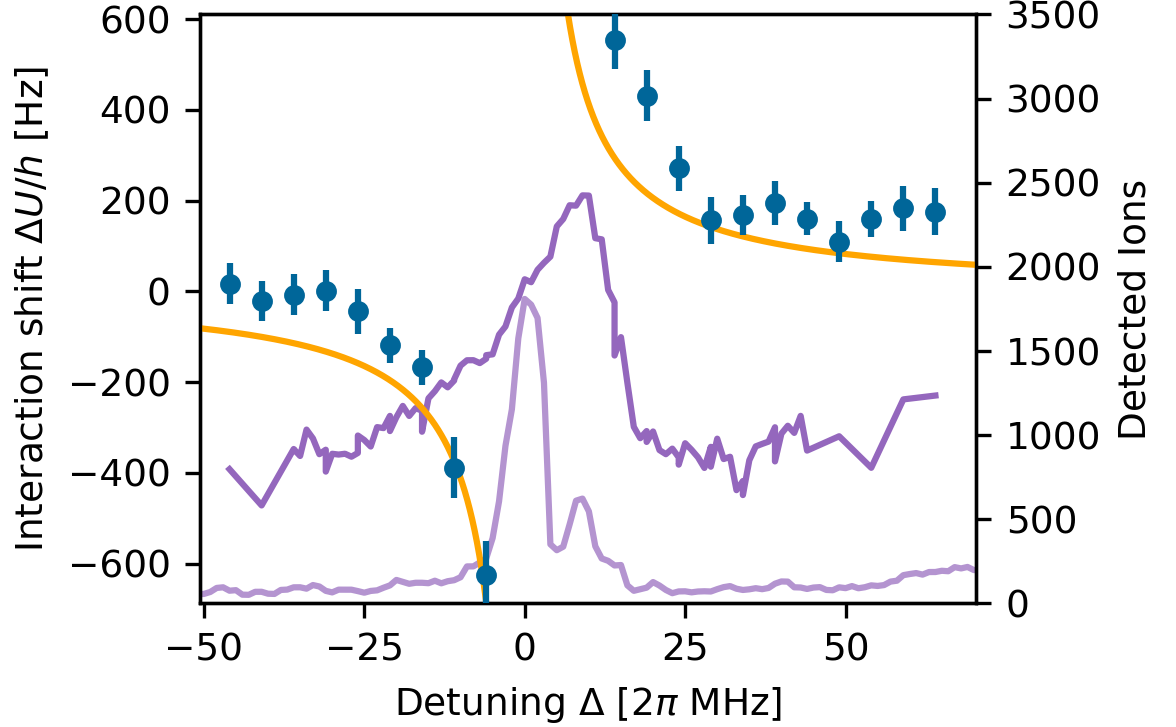}
	\caption{\label{Fig3}Measured interaction shift (blue dots) for different detunings around the photoassociation resonance to the Rydberg molecule shown in Fig.\,\ref{Fig1}, for an atomic Rabi frequency of $2\pi\times\SI{810}{\kHz}$. The measured on-site interaction for the uncoupled system is $U_\mathrm{ref}=h\times\SI{2110\pm20}{\Hz}$, yielding a change in the interaction of up to \SI{30}{\percent}. The shown resonances are the total number of ions detected within the same measurement procedure with a hold time of $\tau = \SI{450}{\us}$ and the same driving strength (purple) and with a reduced Rabi frequency of $2\pi\times\SI{230}{\kHz}$ (light purple). The orange curve is a simplified theoretical model using the experimentally measured atomic Rabi frequency and resonance position (see text). Error bars correspond to statistical errors from the fitting procedure.}
\end{figure} 

The measured interaction shift around the photoassociaton resonance to the Rydberg molecule can be found in Fig.\,\ref{Fig3}. It is clearly visible that the data shows the resonance structure expected for an OFR with increasing (decreasing) interaction for blue (red) detuning. The measured on-site interaction for the uncoupled system is $U_\mathrm{ref}=h\times\SI{2110\pm20}{\Hz}$, leading to a measured change in interaction of up to \SI{30}{\percent}. As the on-site interaction is linearly dependent on the scattering length we can extract the scattering length using
\begin{equation}
	a = \frac{U}{U_\mathrm{ref}} a_{bg}
\end{equation}
with $a_{bg}=\SI{99}{a_0}$ being the background scattering length for rubidium \cite{vanKempen2002scatteringLength}. Our measurements thus correspond to a maximum change in scattering length of $\pm\SI{30}{a_0}$.

From the ion signal for small driving (light purple data in Fig.\,\ref{Fig3}) we can clearly see that the resonance structure splits into two separate lines. We measured the natural lifetime of the two resonances, with a different measurement method (as explained in \cite{Niederpr2015giantCross}), to be \SI{13.2\pm 0.6}{\us} (left peak) and \SI{8.0\pm 0.4}{\us} (right peak). This is of the same order as the lifetime of the underlying atomic Rydberg state. From our theoretical model of the Born-Oppenheimer potentials (see supplementary material) we do not expect any other deeply bound lines in the energetic vicinity of the ground state, so that we conclude that both lines correspond to the ground state in the well at $\approx\SI{700}{a_0}$. The splitting might stem from relativistic effects in the scattering interaction between the electron in the Rydberg state and the bound ground state atom, which we don't take into account in our potential calculations \cite{Eiles2017SpinEffectsRydMol, markson2016SpinEffectsRydMol}. For our theoretical model we ignore the splitting and treat it as an effective single resonance.

As in other experiments \cite{Yan2013OFRintercombSr, Theis2004TuningScatteringOFR, Blatt2011OFRIdealGas} we see strong losses beyond the expected power broadening by the excitation light. While we would expect for our parameters line widths in the \si{kHz} range, we measure resonance structures in the order of several \si{\MHz} with a uniform background of about one third of the height of the resonance peak (purple data in Fig.\,\ref{Fig3}). We attribute this to several points. First of all we are technically limited to a resolution of about \SI{1}{\MHz} due to long term instabilities in the locking procedure and the linewidth of the coupling light. Secondly we believe the large background signal is caused by correlated cluster dynamics stemming from the atomic resonance, which is known to happen in dense driven Rydberg systems \cite{fletscher2017Bistability}. Finally, this interaction induced broadening mechanism should also appear at the molecular line itself. As a result, we observe an overall lifetime of the sample of about \SI{1}{\ms}, which is of similar performance as previous experiments using intercombination lines \cite{Yan2013OFRintercombSr, Blatt2011OFRIdealGas}. However as we will discuss later, this is not a fundamental limit for a Rydberg OFR, but rather due to technical reasons in our particular system.

For the theoretical description of the OFR, we use a 2-level model in the limit of large detuning $\Delta$. The interaction shift is then given by $\Delta U = \hbar\Omega_\mathrm{mol}^2/(4 \Delta)$. We derive the molecular Rabi frequency $\Omega_\mathrm{mol}$ from the atomic one using $\Omega_\mathrm{mol} = \sqrt{2} FC\times \Omega$. The atomic Rabi frequency $\Omega$ is experimentally measured (see supplementary material) and the factor $\sqrt{2}$ accounts for the symmetry of the excited state. The Franck-Condon factor $FC$ is retrieved numerically (see supplementary material). For a lattice depth of $\SI{30}{\erec}$ we obtain $FC = \num{0.12}$ corresponding to a molecular Rabi frequency of $\Omega_\mathrm{mol} = 2\pi\times\SI{140}{kHz}$ for $\Omega = 2\pi\times\SI{810}{\kHz}$. We set the resonance position to be at the center of the stronger molecular peak (Fig.\,\ref{Fig3}). Despite these simplifications, this ab initio theory with no adjustable parameter shows good agreement with the measured data. It can be shown that our treatment as a bound-bound transition is equivalent to the usual treatment as a free-bound transition provided the size of the molecule lies within the linear part of the relative wave function of the two ground state atoms (see supplementary material).

\begin{figure}
	\includegraphics[width=1\columnwidth]{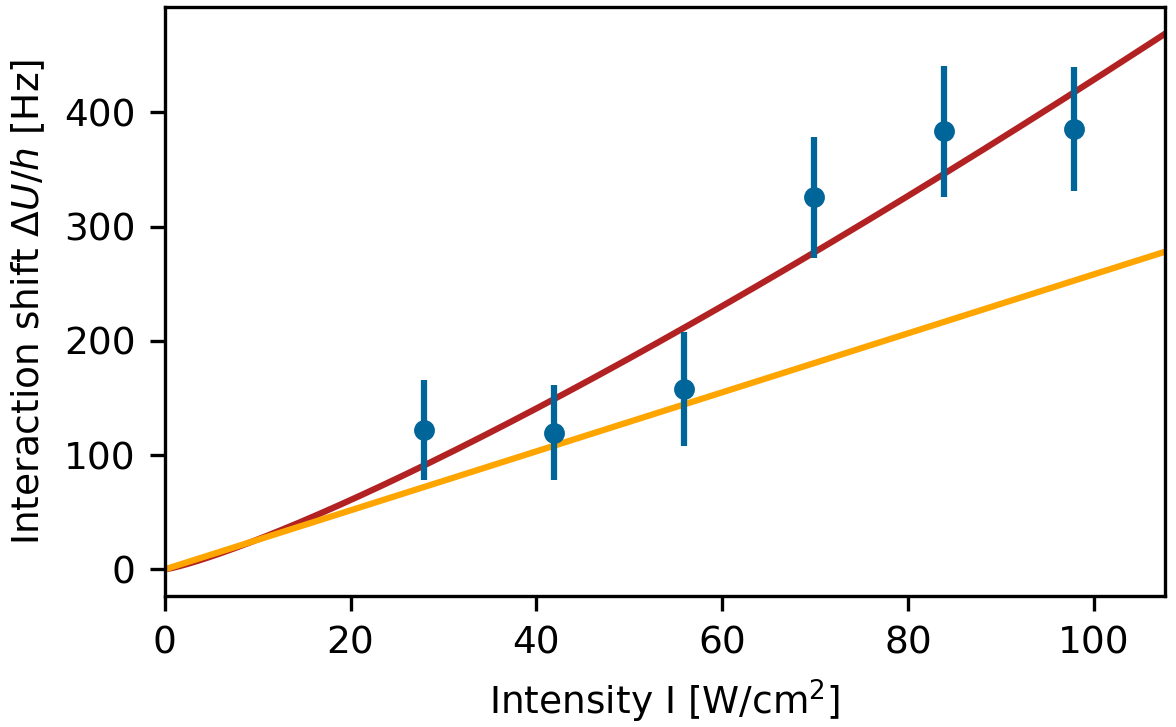}
	\caption{\label{Fig4}Measured interaction shift $\Delta U$, for different coupling laser intensities and a detuning $\Delta = 2\pi\times\SI{+18}{\MHz}$. The data points at $\approx \SI{85}{\W\per\cm\squared}$ correspond to an atomic Rabi frequency of $\Omega = 2\pi\times\SI{810}{\kHz}$ as in Fig.\,\ref{Fig3}. The data fits well to the 2-level model for low intensities (orange), however for strong driving it increases non-linearly. A power law fit (red) yields a scaling with $\propto I^{1.2}$. Error bars correspond to statistical errors from the fitting procedure.}
\end{figure}

An important aspect of a Rydberg OFR is the role of Rydberg-Rydberg interactions. As discussed above, this leads to strongly enhanced losses. In addition, one should also expect that the presence of real atomic and molecular Rydberg excitations in the sample leads to an interaction induced shift of the molecular resonance. Indeed, the $C_6$ coefficient is positive for the involved state and we expect a blue shift of the molecular state. Consequently, the effective detuning is reduced for blue detuning and increased for red detuning. This effect can be seen in Fig.\,\ref{Fig4}, where we show the intensity dependence of the interaction shift for blue detuning. While the ab initio theory can only describe a linear increase, a power law fit to the experimental data reveals an exponent of $\propto I^{1.2}$. We attribute this nonlinearity to the effective blue-shift of the resonance. 

\begin{figure}
	\includegraphics[width=1\columnwidth]{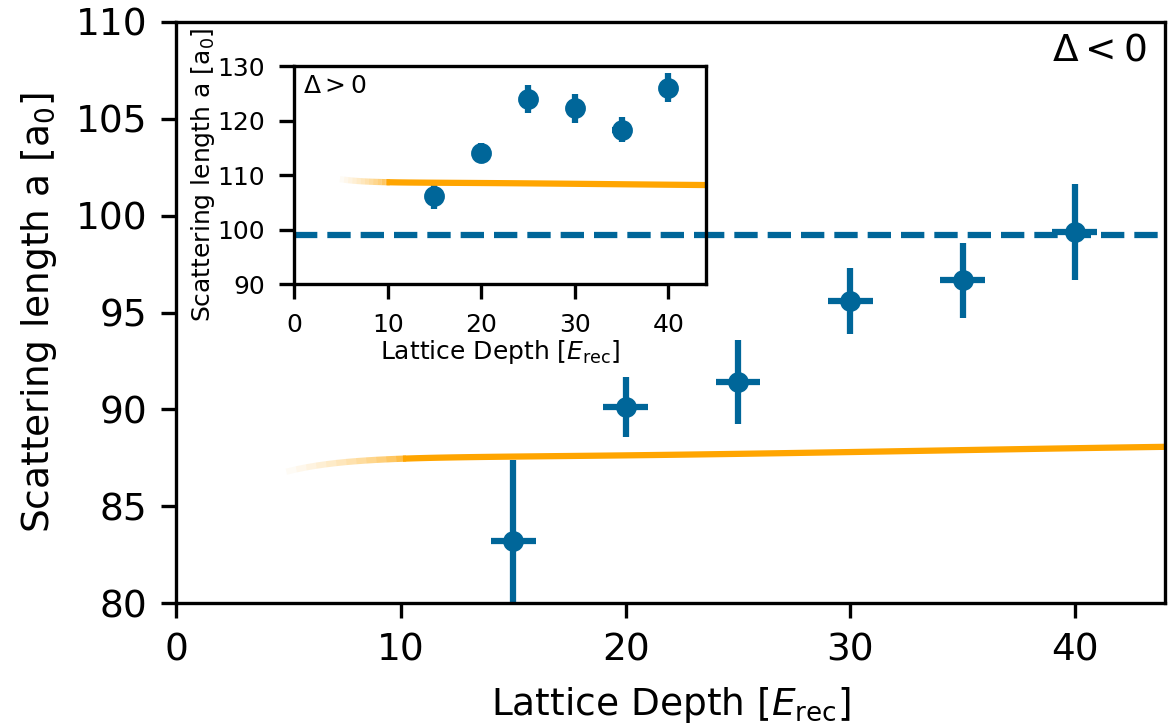}
	\caption{\label{Fig5}Dependence of the measured scattering length $a = U/U_\mathrm{ref}\, a_{bg}$ on the final lattice depth for red detuning $\Delta = 2\pi\times\SI{-17}{\MHz}$ (inset for blue detuning $\Delta = 2\pi\times\SI{18}{\MHz}$). While the calculated theory (orange) shows a near constant behaviour the experimental data shows an increase in scattering length with increasing lattice depth for both blue and red detuning. We attribute this to background Rydberg excitations leading to a shift towards positive detuning for the molecular lines (see text). Error bars in the scattering length correspond to the statistical errors from the fitting procedure. Error bars in the lattice depth (\SI{\pm 1}{\erec}) are an estimated systematic error.}
\end{figure}

Further evidence for this interpretation can be found from varying the lattice depth. An increase in lattice depth causes a larger Franck-Condon overlap, as the Wannier function is compressed. This leads to an increase in the interaction shift. At the same time, the on-site interaction energy $U_\mathrm{ref}$ stemming from the background scattering length scales exactly in the same way, such that the effective scattering length remains unchanged. The lattice depth dependence therefore measures the influence of the Rydberg population on the scattering length. Our experimental data (Fig.\,5) show indeed an increase in scattering length for blue as well as red detuning, while for small lattice depths, the ab initio theory is recovered. This nonlinearity might also explain the previously not discussed variations seen in Fig.\,\ref{Fig3}, where the measured data tends to be above the theoretical model for all detunings.


Rydberg optical Feshbach resonances have a couple of advantages in comparison to previously observed OFR. First, they are readily available in a lot of atomic species. Rydberg molecules have been experimentally measured in rubidium \cite{bendkowsky2009observation}, cesium \cite{Booth2015CsMolecules} and strontium \cite{camargo2016strontiumRydMol} and they are expected to exist for all species or atomic mixtures having a negative electron-ground state scattering length \cite{sandor2017RydOFR}. This is the case, e.g. for all alkali atoms \cite{greene2000longRangeRydMols}. Secondly, the molecular states can be addressed spin-resolved and even two OFR can be active at the same time. Most importantly however we believe a Rydberg OFR is not limited by the performance shown here. Note that our preparation scheme provides rather high atomic densities and the average occupation number in the center of the cloud is estimated to be about \numrange{6}{8} atoms per lattice site. Our experiment is therefore particularly sensitive to density dependent losses. Thus, reducing the atomic density in future experiments to an average filling factor of $\overline{n}=1$ should significantly reduce these losses. Furthermore, the coupling laser intensity could be increased by a factor of ten compared to the measurements reported here, using commercially available high power UV laser systems. Additionally Rydberg molecules feature a plenitude of available states with long lifetimes. For a state bound at $\approx\SI{550}{a_0}$ in the same potential we have already measured a change in scattering length of $\approx\SI{5}{a_0}$, while maintaining a lifetime of \SI{25}{\ms}. The product of these two numbers defines a figure of merit to characterize a resonance. This thus already corresponds to an increase in performance by a factor of five compared to the discussed resonance in this article. We believe an even better performance can be achieved by using states with lower principle quantum number $n$. While these have a reduced lifetime $\propto n^3$, the interaction shift due to a stronger coupling scales as $\propto n^{-3}$. Because of a deeper binding energy $\propto n^{-6}$ and a reduced Van-der-Waals coefficient $\propto n^{11}$ for Rydberg-Rydberg interactions, it should be possible to operate the OFR outside the spectral region of interaction induced broadening. If that was accomplished, Rydberg OFR could become competitive to magnetic Feshbach resonances, with the additional advantage of combining fast temporal modulations of the interaction on short length scales. Coupling to more exotic molecular states such as the recently discovered Butterfly or Trilobite molecules \cite{niederprum2016observation, Booth2015CsMolecules}, possessing permanent electric dipole moments up to several kilodebye, could even be a way to tune anisotropic interactions via the p-wave scattering length.

There is one more aspect, which makes Rydberg OFR truly unique and appealing. This is the availability of molecular trimer or tetramer states \cite{Bendowsky2010Trimer}. Coupling an atomic gas to such a molecule allows to engineer genuine attractive or repulsive three- and four-body interactions in the gas. Such experiments would open up a completely new and so far unexplored territory for interacting many-body systems.

\section{Methods}
\textbf{Experimental procedure.} To perform our experiments we prepare a Bose-Einstein condensate of $\approx\num{80e3}$ $^{87}\mathrm{Rb}$ atoms, spin polarized in the $\ket{5\mathrm{S}_{1/2}, F=2, m_F=2}$ state, in a crossed optical dipole trap with trapping frequencies $\omega_{x,y,z} = 2\pi\times(\num{61},\num{61},\num{51})\si{Hz}$. We adiabatically transfer these atoms into a blue detuned 3D optical lattice with lattice constants $a_{x,y,z} = (374,374,529)\si{nm}$ and lattice depth $S=\SI{8}{\erec}$ with an exponential ramp with time constant \SI{20}{ms}. The lattice depth is then linearly increased within \SI{50}{\us} to a final depth of $S=\SI{30}{\erec}$, while simultaneously switching off the underlying harmonic trapping potential. The coupling light at \SI{297}{\nm} is generated from a frequency doubled dye laser and the stated atomic Rabi frequencies are calibrated from a different set of measurements (see supplementary material). It is linearly polarized parallel to the quantization axis of the system. We ramp it to it's final power within the same \SI{50}{\us} as increasing the lattice depth. After a variable hold time $\tau$, we switch off the lattice potential as well as the coupling laser instantaneously and take a time-of-flight image of the dropping atomic cloud. Additionally we record the time resolved ion signal during illumination using a small electric field ($E \approx \SI{0.2}{\V\per\cm}$). Ions are created from excited Rydberg states mainly through photoionization from the lattice beams at a small rate compared to the natural decay or the coupling strength. 


\textbf{Determination of the on-site interaction.} As we switch off the underlying trapping potential the collapse and revival dynamics are on top of Bloch-Oscillations in the lattice potential. Thus to extract the super-fluid visibility $\mathscr{V}$, for a given hold time $\tau$, we extract the maximum density of the atomic cloud from a time-of-flight image. From this central peak we take the usual approach defining four boxes around the center containing the super-fluid peaks, as well as four boxes turned by 45 degrees with respect to the first four boxes. We define the visibility $\mathscr{V}$ as the pixel sum from the first set of boxes minus the pixel sum of the second set of boxes divided by the pixel sum of all eight. A perfect super fluid state would thus have a visibility equal to one and a completely collapsed state equal to zero. 

To extract the on-site interaction $U$ we fit the measured visibility using:
\begin{equation}
\begin{split}
	\mathscr{V} &\propto \left| \bra{\alpha(t)}\hat a\ket{\alpha(t)} \right|^2 \mathrm{e}^{-\tau/\tau_0}\\
	 &= \left|\sqrt{\overline n}\exp{(\overline n(\mathrm{e}^{i(-U/\hbar\,\tau+\phi)}-1))}\right|^2 \mathrm{e}^{-\tau/\tau_0}.
\end{split}
\end{equation}
The exponential decay factor accounts for any imperfections and losses in the system. $\overline n$ is the average atom number per site and $\phi$ a phase offset imprinted by the finite ramp length to the collapse depth. For all measurements we simultaneously record a reference measurement without coupling to the molecular state to compensate for day to day drifts and alignment imperfections in the lattice beams and extract the interaction shift $\Delta U = U - U_\mathrm{ref}$. 


\section{Acknowledgements}
We would like to thank M. Eiles for helpful discussions and I. Fabrikant for providing the $e^-$-Rb scattering phase-shifts used to calculate the molecular potential. H.O. acknowledges financial support by the DFG within the SFB/TR 49 and the SFB/TR 185. C.L. acknowledges financial support by the DFG within the SFB/TR 185. O.T. acknowledges financial support by the DFG within the SFB/TR 49 and the MAINZ graduate school.

\section{Author Contribution}
O.T., C.L. and T.E. performed the experiments. O.T. analysed the data and prepared the manuscript. H.O. supervised the project. All authors contributed to the data interpretation and manuscript preparation.


\widetext
\clearpage
\begin{center}
\textbf{\large Supplementary Material: Experimental Realization of a Rydberg optical Feshbach resonance in a quantum many-body system}
\end{center}
\setcounter{equation}{0}
\setcounter{figure}{0}
\setcounter{table}{0}
\setcounter{page}{1}
\makeatletter
\renewcommand{\theequation}{S\arabic{equation}}
\renewcommand{\thefigure}{S\arabic{figure}}

\section{Rabi frequency calibration}
\label{sup:RabiFrequency}

To calibrate the Rabi frequency $\Omega$ of the excitation system we investigate the light shift from the excitation light on the $\ket{F=2, m_F=+2}$ ground state. To do so we prepare an atomic cloud in the $\ket{F=1, m_F=+1}$ state and illuminate it with the excitation light blue detuned to the $\ket{F=2, m_F=+2}$ to $\ket{25\mathrm{P}_{3/2}}$ transition while probing the lower transition with a microwave pulse. From the resonance positions of the $\ket{F=2, m_F=+2}$ state for different intensities and detuning $\Delta$ (in the range of \num{50} to \SI{200}{\MHz}) of the excitation light we calibrate the Rabi frequency to the excited state using the shift in a 2-level system for large detunings of $\Delta E = \hbar\frac{\Omega^2}{4\Delta}$.

\section{Molecular states}
To calculate the potential energy curves we follow the procedure from our previous work described in \,\cite{Niederpr2016spinFlip} including s- and p-wave scattering as well as the ground state hyperfine interaction. This procedure yields the potential energy curves (PEC's) in Born Oppenheimer approximation. The bound state energies as well as the nuclei wave functions are then obtained using a shooting method.

\section{Franck-Condon factor}
To calculate the Franck-Condon factor for a transition we approximate the ground state wave function with the ground state of an isotropic harmonic oscillator not taking into account any interaction effects between two atoms in a lattice site and taking the geometric mean of the trapping frequencies in the different lattice axes. We can then analytically write the relative ground state wave function of two atoms as:
\begin{equation}
\begin{split}
	\Psi_g(R, \Theta, \Phi) &= Y_{00}(\Theta, \Phi) \mathscr{F}_g(R)/R \\
	&= \frac{1}{\sqrt{4\pi}} \frac{2}{\pi^{1/4}}\left(\frac{\mu\omega}{\hbar}\right)^{3/4}\mathrm{e}^{-\frac{\mu\omega}{2\hbar}R^2}.
\end{split}
\end{equation}
Using this and the numerically integrated molecular nuclei wave function $\mathscr{F}_\mathrm{mol}(R)$ we evaluate the Franck-Condon factor using
\begin{equation}
	FC = \int dR \mathscr{F}_\mathrm{mol}^*(R) \mathscr{F}_g(R).
\end{equation}

\section{Equivalence of bound-bound and free-bound transitions in the limit of vanishing lattice potential}
\label{sup:equivalenceBoundBoundFreeBound}

\textbf{Bound-Bound:} In a simplified two level bound-bound model we can express the change in scattering length as $\Delta a = \frac{\Delta U}{U_0}a_{bg}$, where $\Delta U$ is the shift through the light field $\Delta U \approx \hbar \frac{\Omega_\mathrm{mol}^2}{4\Delta}$ (for $\Delta\gg\Omega_\mathrm{mol}$), $U_0$ the unperturbed on-site interaction and $a_{bg}$ the background s-wave scattering length. First we evaluate the on-site interaction for a spherical symmetric lattice site in harmonic approximation. We can then express the Wannier function as a Gaussian of the form:
\begin{equation}
	\omega_0(x) = \left( \frac{m\omega}{\pi\hbar}\right)^{1/4} \exp\left(-\frac{m\omega}{2\hbar}x^2\right)
\end{equation}
where $\omega$ is the trapping frequency of the potential. The on-site interaction is then given as
\begin{align}
	U_0 &= \frac{4\pi\hbar^2}{m}a_{bg}\int dxdydz \left|w_0(x)w_0(y)w_0(z)\right|^4\\
		&= \frac{4\pi\hbar^2}{m}a_{bg}\left(\frac{m\omega}{2\pi\hbar}\right)^{3/2}.
\end{align}
To compare the change in scattering length to the free-bound case it is necessary to split the Rabi frequency into several parts. We thus write
\begin{align}
	\Omega_\mathrm{mol} &= \frac{d_\mathrm{mol}E}{\hbar} FC
\end{align}
where $d_\mathrm{mol}$ is the molecular dipole matrix element, $E$ is the electric field strength of the excitation light and $FC$ is the bound-bound Franck-Condon factor. As before the relative ground state wave function is hereby approximated as a Gaussian wave function in relative coordinates
\begin{equation}
	\mathscr{F}_g(R) = \frac{2}{\pi^{1/4}}\left(\frac{\mu\omega}{\hbar}\right)^{3/4}R\exp\left(-\frac{\mu\omega}{2\hbar}R^2\right),
	\label{equ:boundState}
\end{equation}
with $\mu = m/2$ the reduced mass of the two atoms. We ignore any interaction between the two particles leading to a separation of $\approx a_{bg}$ for $R\rightarrow0$, as it is a cumbersome job to take into account, without adding any beneficial insight into the problem. Putting this together we end up with
\begin{equation}
\begin{split}
	\Delta a &= \frac{\Delta U}{U_0} a_{bg}\\
		&\approx \frac{d_\mathrm{mol}^2E^2}{4\Delta\hbar} \frac{m}{4\pi\hbar^2} \left(\frac{2\pi\hbar}{m\omega}\right)^{3/2} \left|\int\mathscr{F}_\mathrm{mol}(R)\mathscr{F}_g(R) dR\right|^2\\
		&= \frac{d_\mathrm{mol}^2E^2\mu}{2\Delta\hbar^3} \left|\int\mathscr{F}_\mathrm{mol}(R)R\exp\left(-\frac{\mu\omega}{2\hbar}R^2\right) dR\right|^2\\
		&\approx \frac{d_\mathrm{mol}^2E^2\mu}{2\Delta\hbar^3} \left|\int\mathscr{F}_\mathrm{mol}(R)R dR\right|^2. \label{equ:boundbound}
\end{split}
\end{equation}
Where in the last step we approximated $\exp\left(-\frac{\mu\omega}{2\hbar}R^2\right) \approx 1$ for $\omega \rightarrow 0$.

\textbf{Free-Bound:} For a free-bound transition on the other hand we adapt the method from ref.\,\cite{Nicholson2015OFR}. Here the interaction shift is given by $\Delta a = l_{opt}\frac{\Delta\gamma}{\Delta^2+\gamma^2/4}$ with $l_{opt}$ the optical coupling strength and $\gamma$ the decay from the excited state. We will evaluate this in the limit $\Delta \gg \gamma$ neglecting the second term in the denominator leading to $\Delta a = l_{opt} \gamma/\Delta$. The optical coupling strength is given by Fermi's golden rule
\begin{equation}
	l_{opt} = \frac{\Gamma(k)}{2k\gamma} = \frac{1}{2k\gamma}\frac{2\pi}{\hbar}\left|\bra{n} V_{opt} \ket{E} \right|^2
\end{equation}
where $\ket{n}$ is the normalized excited state and $\ket{E}$ is an energy normalized free state given by
\begin{equation}
	\braket{R|E} = \sqrt{\frac{2\mu}{\pi\hbar^2k}}\sin(kR).
\end{equation}
We again have ignored any interaction effects for $R\rightarrow0$, and $V_{opt}$ is the optical coupling potential. We rewrite
\begin{equation}
\begin{split}
	\bra{n} V_{opt} \ket{E} &= \frac{\hbar\Omega_\mathrm{mol}}{2} = \frac{d_\mathrm{mol}E}{2} FC\\
		&= \frac{d_\mathrm{mol}E}{2} \sqrt{\frac{2\mu}{\pi\hbar^2k}} \int \mathscr{F}_\mathrm{mol}(R) \sin(kR) dR\\
		&\approx \frac{d_\mathrm{mol}E}{2} \sqrt{\frac{2\mu k}{\pi\hbar^2}} \int \mathscr{F}_\mathrm{mol}(R)RdR
\end{split}
\end{equation}
where the last step is valid for $k\rightarrow0$, as is the case for ultra cold atomic clouds. Putting all this together we end up with
\begin{equation}
\begin{split}
	\Delta a &\approx \frac{1}{2k\Delta} \frac{2\pi}{\hbar}  \frac{d_\mathrm{mol}^2E^2}{4} \frac{2\mu k}{\pi\hbar^2} \left|\int \mathscr{F}_\mathrm{mol}(R)R dR \right|^2\\
		&= \frac{d_\mathrm{mol}^2E^2\mu}{2\Delta\hbar^3} \left|\int\mathscr{F}_\mathrm{mol}(R)R dR\right|^2
\end{split}
\end{equation}
which is equivalent to equation \ref{equ:boundbound} for a bound-bound description.

%

\end{document}